\documentclass{jpp}
\usepackage{epstopdf, epsfig}
\usepackage{hyperref}
\usepackage{amssymb,amsmath,amsfonts,graphicx,epsf}
\usepackage{color}

\newcommand\equr{\mathrel{\overset{\makebox[0pt]{\tiny\mbox{UR}}}{=}}}
\newcommand\eqnr{\mathrel{\overset{\makebox[0pt]{\tiny\mbox{NR}}}{=}}}

\shorttitle{Maximum entropy}
\shortauthor{V. Zhdankin}

\title{Nonthermal particle acceleration from maximum entropy in collisionless plasmas}

\author{Vladimir Zhdankin\aff{1}
  \corresp{\email{vzhdankin@flatironinstitute.org}}} 

\affiliation{\aff{1}Center for Computational Astrophysics, Flatiron Institute, 
162 Fifth Avenue, New York, NY 10010, USA}

\begin{document}

\maketitle

\begin{abstract}
Dissipative processes cause collisionless plasmas in many systems to develop nonthermal particle distributions with broad power-law tails.  The prevalence of power-law energy distributions in space/astrophysical observations and kinetic simulations of systems with a variety of acceleration and trapping (or escape) mechanisms poses a deep mystery. We consider the possibility that such distributions can be modeled from maximum-entropy principles, when accounting for generalizations beyond the Boltzmann-Gibbs entropy. Using a dimensional representation of entropy (related to the Renyi and Tsallis entropies), we derive generalized maximum-entropy distributions with a power-law tail determined by the characteristic energy scale at which irreversible dissipation occurs. By assuming that particles are typically energized by an amount comparable to the free energy (per particle) before equilibrating, we derive a formula for the power-law index as a function of plasma parameters for magnetic dissipation in systems with sufficiently complex topologies. The model reproduces several results from kinetic simulations of relativistic turbulence and magnetic reconnection.
\end{abstract}

\section{Introduction}

Nonthermal energetic particles are ubiquitous in collisionless plasmas, being observed in laboratory experiments \citep[e.g.,][]{yoo_etal_2013, bulanov_etal_2015, schroeder_etal_2021}, planetary magnetospheres \citep{birn_etal_2012}, the solar wind \citep{fisk_gloeckler_2007}, the solar corona \citep{aschwanden_2002}, and high-energy astrophysical systems \citep[e.g.,][]{blandford_eichler_1987}. It was long recognized that nonthermal particles are a generic consequence of collisionless plasma physics, as the absence of Coulomb collisions precludes relaxation to a thermal equilibrium \citep[e.g.,][]{fermi_1949, fermi_1954, parker_tidman_1958}. More recently, first-principles numerical simulations demonstrated efficient particle acceleration from shocks \citep{spitkovsky_2008, sironi_spitkovsky_2010, caprioli_spitkovsky_2014}, magnetic reconnection \citep{sironi_spitkovsky_2014, guo_etal_2014, werner_etal_2016, li_etal_2019}, relativistic turbulence \citep{zhdankin_etal_2017, comisso_sironi_2018}, and various instabilities \citep[e.g.,][]{hoshino_2013, kunz_stone_quataert_2016, nalewajko_etal_2016, alves_etal_2018, ley_etal_2019, sironi_etal_2021}. In observations and simulations, particle energy distributions frequently exhibit power-law tails in which the index $\alpha$ can range from hard ($\alpha \sim 1$) to soft ($\alpha \gg 1$) values, depending on system parameters. Determining why power-law distributions form and predicting $\alpha$ as a function of parameters are topics of fundamental importance.

This Letter explores the possibility that power-law distributions in collisionless plasmas can be explained by maximum-entropy principles, when considering nonextensive entropy measures beyond the traditional Boltzmann-Gibbs (BG) entropy. There is no {\it a priori} reason for a collisionless plasma to relax to a state of maximum BG entropy. Given that plasma dissipation processes are macroscopically irreversible, the question is then, what type of entropy (if any) does a collisionless plasma maximize upon equilibration?

Generalized measures of entropy form a possible foundation for non-equilibrium statistical mechanics. In particular, the non-extensive entropy of \cite{tsallis_1988}, building on earlier ideas by \cite{renyi_1961} and others, has gained attention in the community. Non-extensive entropy was suggested to be relevant for physical systems with long-range correlations \citep[e.g.,][]{milovanov_zelenyi_2000}, which are a generic outcome of nonlinear processes in collisionless plasma. It was shown that the maximization of Tsallis entropy leads to the kappa distribution, which has a quasi-thermal peak along with a power-law tail that extends to high energies \citep{milovanov_zelenyi_2000, leubner_2002, livadiotis_mccomas_2009}. Incidentally, the kappa distribution is widely used to model nonthermal particle distributions in space plasmas such as the solar wind \citep[e.g.,][]{pierrard_lazar_2010, livadiotis_mccomas_2013}. While intriguing, the generalized measures of entropy have degrees of freedom (e.g., the entropic index or kappa index) that are not straightforward to interpret physically or model phenomenologically, which has limited their utility.

Recently, \cite{zhdankin_arxiv} developed a framework for quantifying generalized entropy based on dimensional representations of entropy, derived from the Casimir invariants of the Vlasov equation. This framework shares similarities to the non-extensive entropies of \cite{renyi_1961} and \cite{tsallis_1988}, but enables a connection with irreversible processes occurring at various energy scales within the plasma. Thus, long-range correlations are re-interpreted as the relaxation of a collisionless plasma subject to dissipation at nonthermal energies. In this Letter, we use this framework to derive a generalized maximum-entropy (GME) distribution (equivalent to the Tsallis distribution) that has a power-law tail at high energies, resembling numerical and observational results in the literature. For a given number of particles and kinetic energy content, there is only one unconstrained free parameter (linked to $\alpha$), determined by the energy scale at which entropy is maximized.

After deriving the GME distribution, we propose a model for determining the power-law index $\alpha$ as a function of physical parameters, for systems governed by magnetic dissipation with sufficiently complex topologies. By assuming that particles are typically energized by an amount comparable to the free energy per particle before equilibrating, we derive an equation for $\alpha$ versus plasma beta and fluctuation amplitude, indicating that nonthermal particle acceleration is efficient when $\beta$ is low and fluctuations are strong. We compare the model predictions to numerical results from the literature on relativistic turbulence and magnetic reconnection, showing that the model is able to reproduce some observed trends such as the scaling of $\alpha$ with the magnetization $\sigma$. The GME model also provides a resolution for why power-law distributions are often similar for distinct processes (with diverse escape/trapping mechanisms) and for varying spatial dimensionality (2D versus 3D).

The GME framework provides a route to understanding particle acceleration that is distinct from standard approaches based on quasilinear theory and its extensions. The limitations and applicability of the model are further discussed in the conclusions. 

\section{Model for generalized maximum-entropy distribution}

Consider a collisionless plasma in a closed system. The evolution of the fine-grained particle distribution for a given species can be represented by the (relativistic) Vlasov equation,
\begin{align}
\partial_t f + \boldsymbol{v}\cdot\nabla f + \boldsymbol{F} \cdot \partial_{\boldsymbol{p}} f = 0 \, , \label{eq:vlasov}
\end{align}
where $f(\boldsymbol{x},\boldsymbol{p},t)$ is the particle momentum distribution function (normalized such that $\int d^3p d^3x f = N$ is the total number of particles),~$\boldsymbol{v} = \boldsymbol{p}c/ (m^2c^2+p^2)^{1/2}$ is the particle velocity (with $m$ the particle mass), and~$\boldsymbol{F}(\boldsymbol{x},\boldsymbol{p},t)$ is a phase-space conserving force field ($\partial_{\boldsymbol{p}} \cdot \boldsymbol{F} = 0$), containing the electromagnetic force and external forces. Eq.~\ref{eq:vlasov} can be applied to any particle species, with appropriate $\boldsymbol{F}$. We denote particle kinetic energy by~$E(p) = (m^2c^4+p^2c^2)^{1/2} - m c^2$ and the system-averaged kinetic energy by $\overline{E}$.

The Vlasov equation formally conserves the BG entropy $S=-\int d^3x d^3p f \log{f}$ as well as an infinite set of quantities known as the Casimir invariants. The latter can be manipulated to yield quantities with dimensions of momentum, introduced in \cite{zhdankin_arxiv} as the {\it Casimir momenta}:
\begin{align}
p_{c,\chi}(f) \equiv n_0^{1/3} \left( \frac{1}{N} \int d^3x d^3p f^\chi  \right)^{-1/3(\chi-1)} \, , \label{eq:pchar}
\end{align}
where $n_0$ is the mean particle number density and $\chi > 0$ is a free index that parameterizes the weight toward different regions of phase space: large (small) values of $\chi$ are sensitive to low (high) energies. The phase-space integral in Eq.~\ref{eq:pchar} resembles those used in the nonextensive entropies of \cite{renyi_1961} and \cite{tsallis_1988}. The Casimir momenta, however, manipulate this integral into a dimensional form that is interpretable physically. In particular, the anomalous growth of $p_{c,\chi}$ is indicative of irreversible entropy production at the corresponding momentum scale in phase space (with $\chi \to 0$ corresponding to momenta far in the tail, and $\chi \to \infty$ corresponding to momenta near the mode).

As described in \cite{zhdankin_arxiv}, $p_{c,\chi}$ share many properties with the BG entropy $S$: 1) they reduce to a dimensionalized version of the BG entropy when $\chi \to 1$, as $p_{c,\chi\to1} = n_0^{1/3} e^{S/3N}$; 2) they are maximized when $f$ is isotropic and spatially uniform; and 3) while ideally conserved by the Vlasov equation, the formation of fine-scale structure breaks conservation of $p_{c,\chi}$ for $f$ measured at coarse-grained scales. \cite{zhdankin_arxiv} also argued that $p_{c,\chi}$ associated with coarse-grained $f$ will tend to increase (irreversibly) when energy is injected into the system, for generic complex processes; this was demonstrated by 2D kinetic simulations of relativistic turbulence. Phenomena such as the entropy cascade may lead to anomalous entropy production through finite collisionality \citep{schekochihin_etal_2009, eyink_2018}.

The infinite number of generalized entropies represented by $p_{c,\chi}$ complicates the application of a maximum entropy principle. Only when dissipation occurs collisionally or at small enough energy scales ($\chi \sim 1$) is the BG entropy maximized. In general, mechanisms of anomalous entropy production can operate over a spectrum of scales, so a scale-by-scale understanding of the plasma physical processes is necessary to model the system.

In this Letter, we consider the idealized situation where entropy is maximized at a characteristic momentum scale represented by $p_{c,\chi_d}$ with a given index $\chi_d$ where the subscript $d$ denotes ``dissipation''. Physically, particles are energized up to this scale (on average) while mixing causes the distribution to smooth out to the equilibrium state.

Suppose that the system evolves to maximize $p_{c,\chi_d}$. The GME distribution is isotropic and spatially uniform $f(\boldsymbol{p},\boldsymbol{x})=f(p)$, and can be derived from the functional
\begin{align}
{\mathcal L} = N^{1/3} \left( \int d^3p f^{\chi_d} / N \right)^{-1/3(\chi_d-1)} - \lambda_1 \left( \int d^3p f - N \right) -  \lambda_2 \left[ \int d^3p E(p) f - N \overline{E}\right] \, ,
\end{align}
where $\lambda_i$ are Lagrange multipliers enforcing number and energy constraints. By requiring $\delta {\mathcal L} = 0$ upon variations of the distribution $\delta f$, we obtain
\begin{align}
\frac{p_{c,\chi_d}^{3\chi_d-2} \chi_d f^{\chi_d-1}}{3(1-\chi_d) N^{2/3}} - \lambda_1  -  \lambda_2 E(p) = 0 \,
\end{align}
which leads to the GME distribution
\begin{align}
f = C \left[ E(p)/E_b+ 1 \right]^{-1/(1-\chi_d)} \, , \label{eq:maxentropy}
\end{align}
where $C$ and $E_b$ are the normalization factor and characteristic energy, determined by requiring $4\pi\int dp p^2 f = N$ and $4\pi\int dp p^2 E(p) f = N \overline{E}$. Note that Eq.~\ref{eq:maxentropy} is operationally equivalent to the Tsallis distribution \citep{tsallis_1988}; this equivalence is due to the fact that the Tsallis entropy and Casimir momenta are both obtained from the same fundamental phase-space integral (involving powers of $f$). We will restrict our attention to $\chi_d < 1$, in which case there is a power-law tail (whereas $\chi_d > 1$ would lead to a narrow distribution with sharp cutoff). The derivation of Eq.~\ref{eq:maxentropy} from maximizing a dimensional representation of generalized entropy is the first main result of this work.

In the ultra-relativistic (UR) limit, $\overline{E} \gg m c^2$, the GME distribution (Eq.~\ref{eq:maxentropy}) becomes
\begin{align}
f \xrightarrow[]{\text{UR}} C (p/p_b + 1)^{-\alpha-2} \label{eq:fur}
\end{align}
where $\alpha = (2\chi_d-1)/(1-\chi_d)$, $C = N (\alpha - 1) \alpha (\alpha + 1)/8\pi p_b^3$, and $p_b = (\alpha - 2) \overline{E}/3 c$. In the non-relativistic (NR) limit, $\overline{E} \ll m c^2$, Eq.~\ref{eq:maxentropy} becomes
\begin{align}
f \xrightarrow[]{\text{NR}} C \left( p^2/p_b^2 + 1 \right)^{-\alpha-1/2}  \label{eq:fnr}
\end{align}
where $\alpha = (1+\chi_d)/2(1-\chi_d)$, $C = N \Gamma(\alpha+1/2)/ \pi^{3/2} p_b^3 \Gamma(\alpha-1)$, and $p_b = [4 (\alpha-2) m \overline{E}/3]^{1/2}$. The NR expression (Eq.~\ref{eq:fnr}) is equivalent to the kappa distribution.

In both limits, we used $\alpha$ to denote the power-law index of the corresponding energy distribution,
\begin{align}
F(E) = \frac{dp}{dE} 4\pi p^2 f(p) |_{p=[E (E + 2 m c^2)]^{1/2}/c} \, ,
\end{align}
such that $F(E) \propto E^{-\alpha}$ at high energies. Also note that $\chi_d \to 1$ ($\alpha \to \infty$) recovers the thermal (Maxwell-J\"{u}ttner) distribution, using the identity $(A/x + 1)^{-x} = e^{-A}$ as $x \to \infty$ for any $A$.

Since the GME distribution has an infinite extent in energy, $\alpha > 2$ is necessary for finite $\overline{E}$. Thus, the domain is $3/4 < \chi_d < 1$ for the UR case and $3/5 < \chi_d < 1$ for the NR case. We note that the GME framework can be extended to allow $1 < \alpha < 2$ if an additional constraint is imposed to make the distribution vanish at a maximum momentum $p_{\rm max}$ (which may be related to the system confinement scale, for example). This would be implemented by adding a third Lagrange multiplier to ${\mathcal L}$ that enforces $p_{c,\chi\to 0} = p_{\rm max}$. However, the resulting equation does not have an analytically tractable solution for $f$, so we defer such an extension to future work. This extension may be necessary to accurately model particle distributions in relativistic magnetic reconnection \citep{sironi_spitkovsky_2014} or turbulence \citep{zhdankin_etal_2017} at high magnetization, where $\alpha < 2$ has been measured in simulations.

\section{Model for power-law index}

\begin{figure}
  \includegraphics[width=\columnwidth]{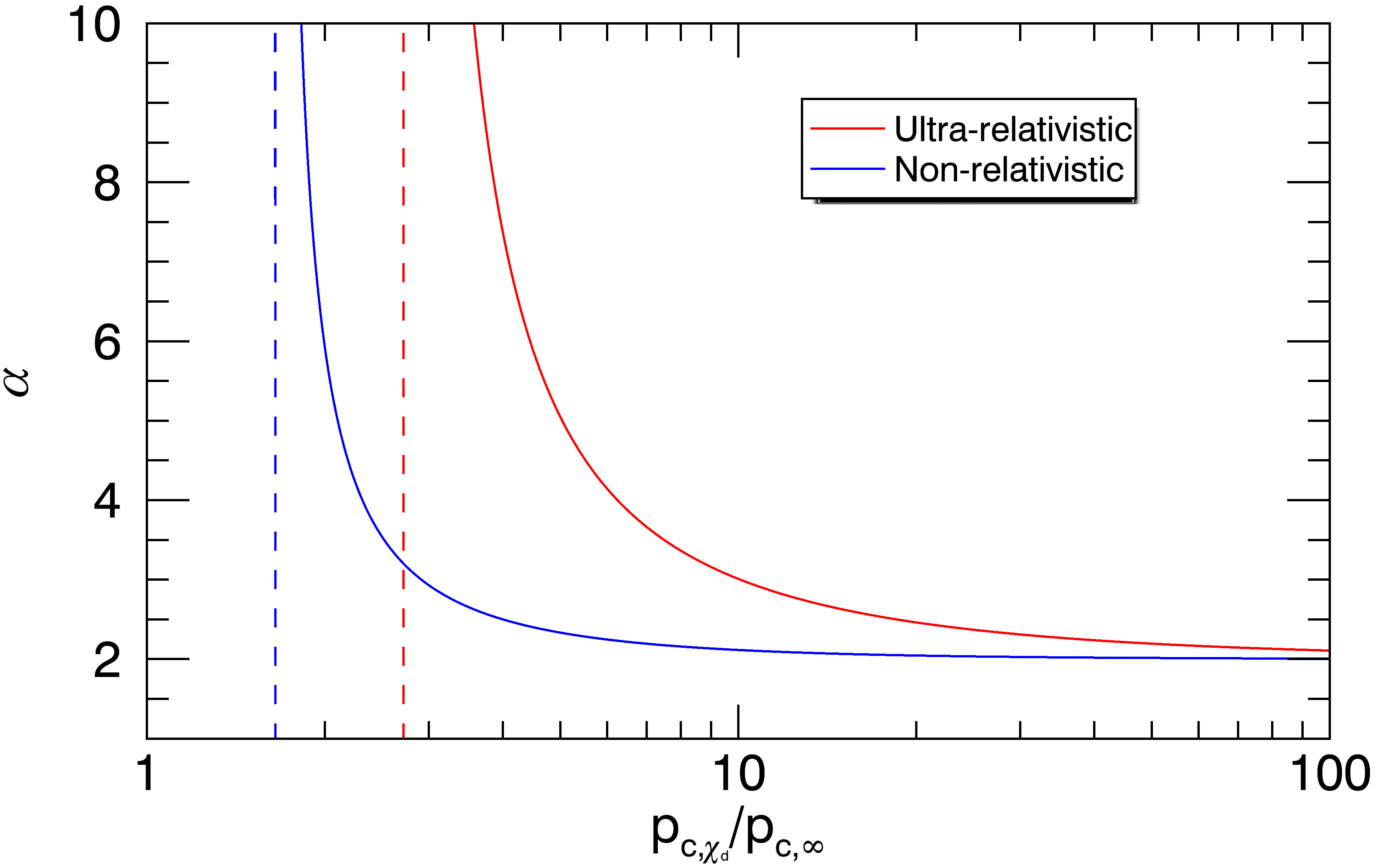} 
   \centering
   \caption{\label{fig:alphatheory} The energy power-law index $\alpha$ of the GME distribution versus the ratio between the entropy-maximizing momentum $p_{c,\chi_d}$ and the typical momentum $p_{c,\infty}$ (Eqs.~\ref{eq:pratio_ur}-\ref{eq:pratio_nr}). The UR (red) and NR (blue) limits are shown separately, with dashed lines indicating singularities.}
 \end{figure}

Suppose that the dynamics are sufficiently complex to cause the initial distribution (which is arbitrary) to evolve into the GME state. One can then compare the momentum at which entropy is maximized, $p_{c,\chi_d}$, with the momentum of the typical particle given by $p_{c,\infty}$ (note that $p_{c,\infty}$ lies close to $p_b$). Evaluating $p_{c,\chi_d}/p_{c,\infty}$ using Eq.~\ref{eq:pchar} with the GME distribution (Eq.~\ref{eq:maxentropy}), one obtains in the UR limit:
\begin{align}
\frac{p_{c,\chi_d} }{p_{c,\infty}} \xrightarrow[]{\text{UR}} \left(\frac{\alpha+1}{\alpha-2} \right)^{(\alpha+2)/3} \, , \label{eq:pratio_ur}
\end{align}
and in the NR limit:
\begin{align}
\frac{p_{c,\chi_d} }{p_{c,\infty}} \xrightarrow[]{\text{NR}} \left( \frac{\alpha-1/2}{\alpha-2} \right)^{(2\alpha+1)/6} \, . \label{eq:pratio_nr}
\end{align}
This relates the power-law index $\alpha$ to the maximum-entropy scale, which can be modeled phenomenologically (as considered below). In Fig.~\ref{fig:alphatheory}, we show $\alpha$ versus $p_{c,\chi_d}/p_{c,\infty}$, separately for the UR (Eq.~\ref{eq:pratio_ur}) and NR (Eq.~\ref{eq:pratio_nr}) limits. Note the divergence $\alpha \to \infty$ when $p_{c,\chi_d}/p_{c,\infty} \to e \approx 2.72$ (UR case) or $p_{c,\chi_d}/p_{c,\infty} \to e^{1/2} \approx 1.65$ (NR case). Thus, if entropy is maximized at momentum scales sufficiently close to the peak of the distribution, then a thermal distribution is recovered (similar to a collisional plasma). When $p_{c,\chi_d}/p_{c,\infty}$ becomes larger than a factor of few, the nonthermal state is obtained, with $\alpha \to 2$ for $p_{c,\chi_d}/p_{c,\infty} \gg 1$. Thus, in both the UR and the NR limit, {\it the distribution will relax to the nonthermal state if entropy is maximized at momenta scales in the tail of the distribution}.

Physical considerations are necessary to determine $p_{c,\chi_d}/p_{c,\infty}$ as a function of system parameters, from which one can extract $\alpha$. In general, this will need to be informed by numerical simulations and analytical considerations for the given process. 
 
\begin{figure}
  \includegraphics[width=\columnwidth]{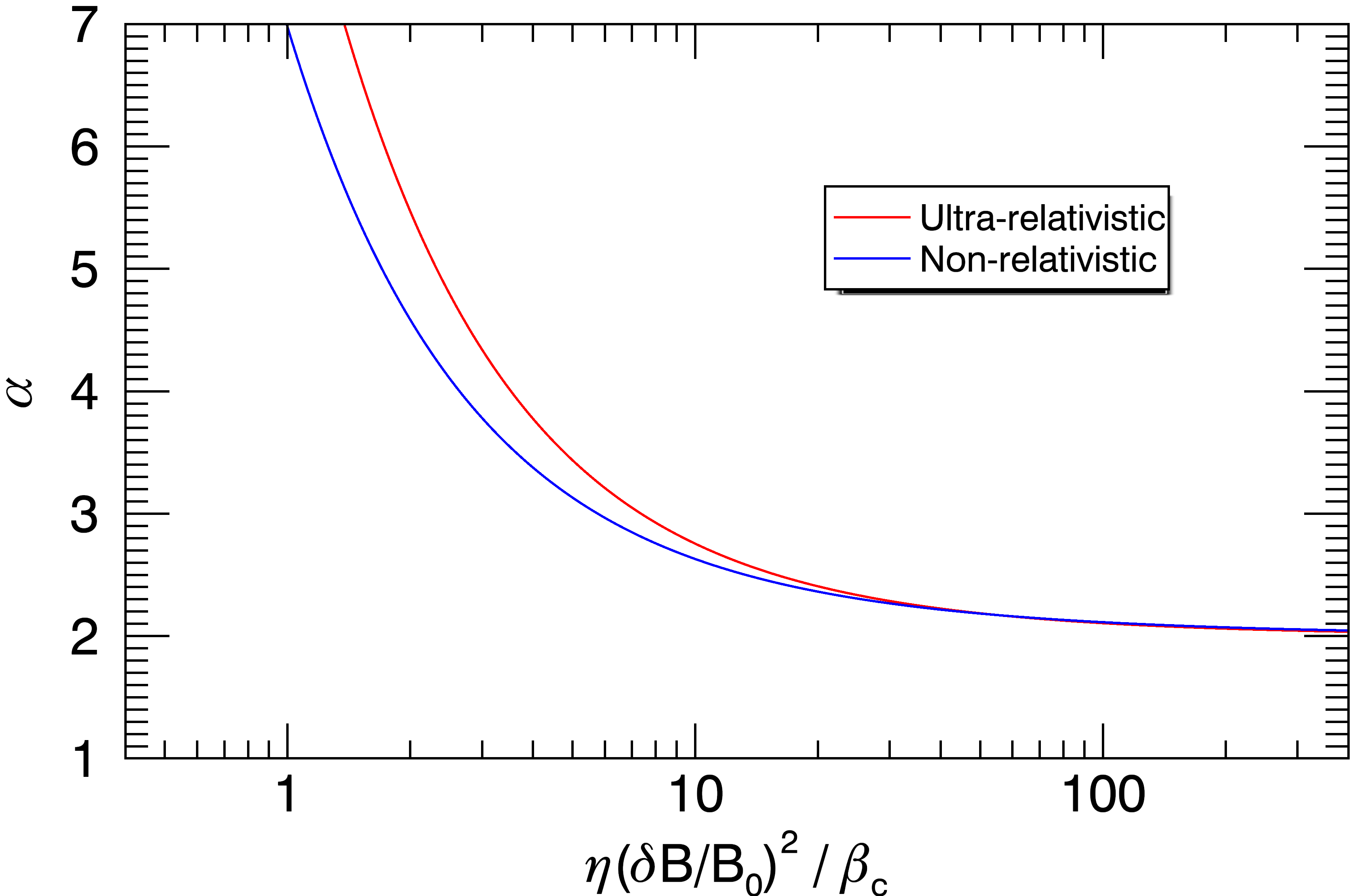} 
   \centering
   \caption{\label{fig:alphabeta} The energy power-law index $\alpha$ of the GME distribution versus (pre-dissipation) physical parameters $\eta (\delta B/B_0)^2 / \beta_c$ for the magnetic dissipation model. The UR (red; Eq.~\ref{eq:alphaur}) and NR (blue; Eq.~\ref{eq:alphanr}) limits are shown separately.}
 \end{figure}

For this Letter, we consider a simplified scenario to estimate the momentum scale of maximum entropy that arises from the dissipation of magnetic energy in complex field topologies (via magnetic reconnection, turbulence, or instabilities). We suppose that rather than being energized at the thermal energy scale, the typical particles are energized by an amount comparable to the free magnetic energy per particle, $E_{\rm free} = \delta B^2/8\pi n_0$, over a dynamical timescale, before equilibrating to the GME state. Here, $\delta B$ is the characteristic magnetic field fluctuation (prior to dissipation), while we denote the background (non-dissipating) component by $B_0$. We denote the energy corresponding to the Casimir momenta by $E_{c,{\chi}} = E(p_{c,\chi})$ and the typical particle energy as $E_0$ (prior to dissipation), noting that the thermal dissipation energy scale is $e E_0$. The model posits that $E_{c,\chi_d} \sim e E_0 + \eta E_{\rm free}$ where $\eta$ is an order-unity coefficient describing the portion of free energy converted. We can then write
\begin{align}
\frac{p_{c,\chi_d}}{p_{c,\infty}} &= \left[ \frac{E_{c,\chi_d} ( E_{c,\chi_d} + 2 m c^2 )}{E_{c,\infty} ( E_{c,\infty} + 2 m c^2)} \right]^{1/2} \nonumber \\
&\sim \left[ \frac{(e E_0 + \eta E_{\rm free}) (e E_0 + \eta E_{\rm free}+2 m c^2)}{E_0 (E_0+2 m c^2)} \right]^{1/2} \nonumber \\
&\sim \left[ \frac{[e + \eta (\delta B/B_0)^2/\beta_c] [e+ \eta (\delta B/B_0)^2/\beta_c+2/\theta_c]}{1+2/\theta_c} \right]^{1/2}\, , \label{eq:master}
\end{align}
where $\theta_c = E_0/m c^2$ is a characteristic dimensionless temperature and $\beta_c = 8\pi n_0 E_0/B_0^2$ is a characteristic plasma beta for the particle species (which may differ from the standard plasma beta, $\beta_0=8\pi n_0 T/B_0^2$ where $T$ is species temperature, by a factor of order unity). Equating Eq.~\ref{eq:master} with either Eq.~\ref{eq:pratio_ur} or Eq.~\ref{eq:pratio_nr} yields an implicit equation for $\alpha$ as a function of $\beta_c$, $\delta B/B_0$, and $\theta_c$ in the appropriate limit. The physical parameters required to achieve a given value of $\alpha$ can then be expressed in the UR limit ($\theta_c \gg 1$) as:
\begin{align}
\eta \left( \frac{\delta B}{B_0} \right)^2 \frac{1}{\beta_c}  \equr \left(\frac{\alpha+1}{\alpha-2} \right)^{(\alpha+2)/3} -e \, , \label{eq:alphaur}
\end{align}
and in the NR limit ($\theta_c \ll 1$) as:
\begin{align}
\eta \left( \frac{\delta B}{B_0} \right)^2 \frac{1}{\beta_c}  \eqnr \left( \frac{\alpha-1/2}{\alpha-2} \right)^{(2\alpha+1)/3} - e \, . \label{eq:alphanr}
\end{align}
The predicted scaling of $\alpha$ given by Eqs.~\ref{eq:alphaur}-\ref{eq:alphanr} is the second main result of this work. The right hand side of both equations becomes zero when $\alpha \to \infty$, indicating that the thermal distribution is recovered for high beta or weak fluctuations, $(\delta B/B_0)^2 / \beta_c \ll 1$. On the other hand, the nonthermal state is obtained when $(\delta B/B_0)^2/\beta_c \gtrsim 1$, for both UR and NR regimes. For $(\delta B/B_0)^2/\beta_c \gg 1$, the index approaches $\alpha \to 2$ (but recall that the model may be extended, in principle, to allow $1 < \alpha < 2$). The scaling is plotted in Fig.~\ref{fig:alphabeta}.

 \section{Comparison to simulations}
 
 To validate the GME model, we remark on how the predictions compare to existing results from kinetic simulations of relativistic turbulence and magnetic reconnection in the literature.

\begin{figure}
 \includegraphics[width=\columnwidth]{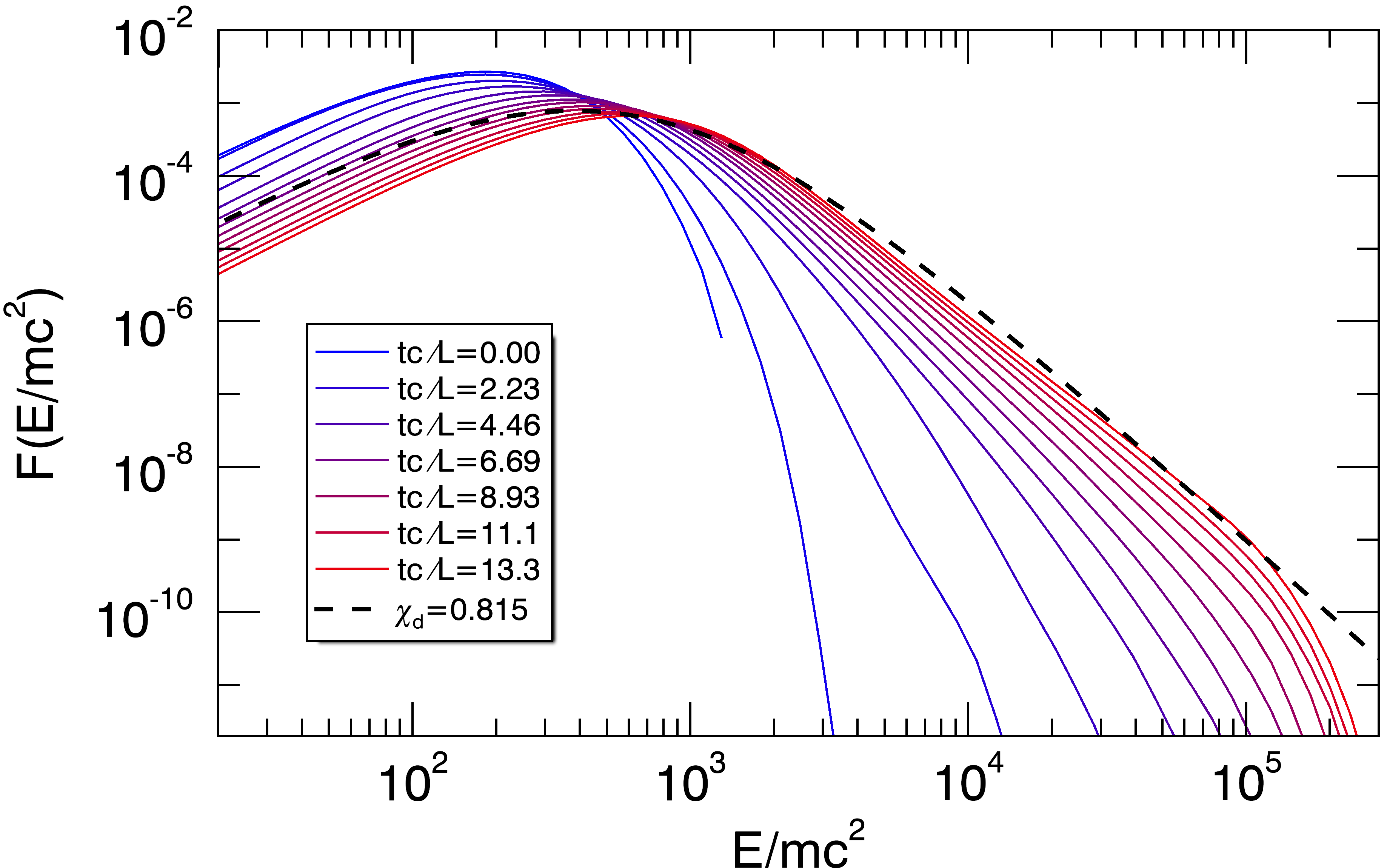}
   \centering
   \caption{\label{fig:distsim} Energy distribution $F(E)$ in PIC simulation of relativistic turbulence for various times, taken from \cite{zhdankin_etal_2018b}, compared to the GME distribution (dashed; Eq.~\ref{eq:fur}) with $\chi_d = 0.815$.}
 \end{figure}

In Fig.~\ref{fig:distsim}, we show the global particle energy distribution $F(E)$ arising in a $1536^3$-cell particle-in-cell (PIC) simulation of driven relativistic turbulence (with $\delta B/B_0 \approx 1$) studied in Refs.~\citep{zhdankin_etal_2018b, wong_etal_2020}. The simulation begins with a Maxwell-J\"{u}ttner distribution of electrons and positrons with UR temperature $\theta=T/m_e c^2=100$ and initial magnetization $\sigma_0 = 3/8$. The magnetization is defined as the ratio of the magnetic enthalpy to plasma enthalpy, and is related to species plasma beta by $\sigma_0 = 1/(4\beta_0)$ in the UR regime; thus $\beta_0 = 2/3$. The simulation develops a nonthermal tail with index $\alpha \approx 3$. We find that the GME distribution of Eq.~\ref{eq:fur} provides a fair fit to the fully developed state when we choose $\chi_d = 0.815$, as shown by the dashed line in Fig.~\ref{fig:distsim}. The fit over-predicts the number of particles at energies below the peak, indicating that relaxation to the GME state is incomplete (possible reasons for this will be described in the conclusions). The PIC simulations of decaying, magnetically-dominated turbulence by \cite{comisso_sironi_2019} also appear to resemble the GME state. Thus, we believe that the GME model provides a reasonable (if imperfect) representation of available numerical data on relativistic turbulence.

\begin{figure}
\includegraphics[width=\columnwidth]{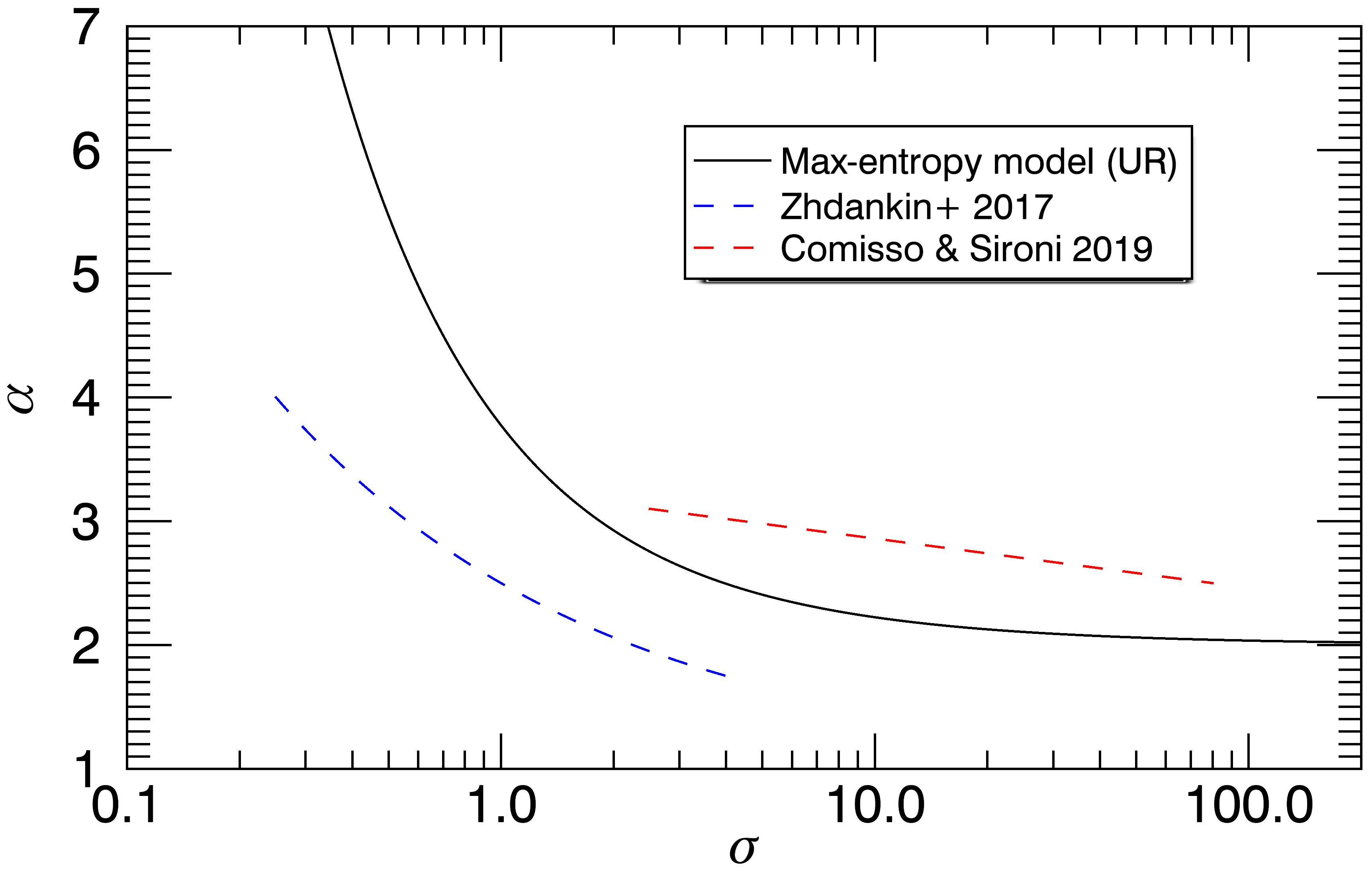} 
   \centering
   \caption{\label{fig:alphasim} Energy power-law index $\alpha$ versus magnetization $\sigma$ from the GME model in the UR limit (black; Eq.~\ref{eq:alphaur} with $\beta_c = 1/4\sigma$, $\delta B/B_0 = 1$, and $\eta = 1$) compared to empirical fitting formula $\alpha \approx \alpha_\infty + C_0 \sigma^{-0.5}$ from PIC simulations of driven relativistic turbulence in \cite{zhdankin_etal_2017} (blue). Also shown is the approximate range of indices from PIC simulations of decaying relativistic turbulence from \cite{comisso_sironi_2019} (their Fig. 5 inset; red).}
 \end{figure}

We next consider the model for the power-law index $\alpha$ from magnetic dissipation. In Fig.~\ref{fig:alphasim}, we compare the predicted $\alpha$ versus $\sigma$ scaling (Eq.~\ref{eq:alphaur} with $\beta_c = 1/4\sigma$, $\delta B/B_0 = 1$, $\eta = 1$) with results in the literature on relativistic turbulence in pair plasma. PIC simulations of driven relativistic turbulence indicate that the power-law index is well-described by the empirical formula $\alpha \approx \alpha_\infty + C_0 \sigma^{-0.5}$, with $\alpha_\infty \approx 1$ and $C_0 \approx 1.5$ for large sizes \citep{zhdankin_etal_2017, zhdankin_etal_2018b}, shown in Fig.~\ref{fig:alphasim} (blue); note that a similar formula with different coefficients was also suggested for relativistic magnetic reconnection \citep{werner_etal_2018, ball_etal_2018, uzdensky_2022}. We also show approximate data points from the 2D decaying relativistic turbulence simulations of \cite{comisso_sironi_2019} (red). The model is able to explain the trends in the numerical simulations fairly well, up to a factor of order unity in $\sigma$. Fits to the simulation data can be improved by adjusting $\eta$, noting that driven turbulence would effectively have a larger $\eta$ than decaying turbulence. Additionally, we note that \cite{comisso_sironi_2019} finds that $\alpha$ increases with decreasing $\delta B/B_0$, consistent with the GME prediction.

In addition to these quantitative comparisons, the GME model provides a resolution to several mysterious findings from kinetic simulations in the literature. Kinetic simulations of disparate processes (turbulence, magnetic reconnection, and instabilities) often exhibit very similar power-law distributions for given plasma parameters. For example, PIC simulations find comparable nonthermal particle acceleration from magnetic dissipation with different current sheet geometries and ensuing dynamics \citep[e.g.,][]{werner_uzdensky_2021}. PIC simulations of relativistic turbulence find that nonthermal particle distributions have a similar shape for different driving mechanisms (electromagnetic, solenoidal, compressive, imbalanced), despite different timescales to arrive at those distributions \citep{zhdankin_2021, hankla_etal_2022}. The universality revealed by these findings may be explained by all of the processes having sufficient complexity to attain a GME state at similar energy scales.

Kinetic simulations also indicate that nonthermal particle distributions formed by relativistic magnetic reconnection \citep{werner_uzdensky_2017, guo_etal_2021} and turbulence \citep{comisso_sironi_2019} are insensitive to the number of spatial dimensions (2D vs 3D), despite different secondary instabilities, cascades, and trapping mechanisms (e.g., long-lived plasmoids in 2D). The GME framework predicts that the distributions are insensitive to the number of spatial dimensions, as long as there are sufficient degrees of freedom to attain such a state.

The GME model predicts similar acceleration efficiency in the NR regime as in the UR regime, as long as no factors arise that suppress entropy production at high energies. PIC simulations in the NR regime are generally constrained in scale separation, which may limit power-law formation. However, recent PIC simulations of NR magnetic reconnection provide some evidence for (steep) power-law distributions at low $\beta$ \citep{li_etal_2019}. Recent simulations of reduced kinetic models indicate efficient electron acceleration by NR magnetic reconnection at macroscopic scales when $\delta B/B_0$ is large enough \citep{arnold_etal_2021}. Hybrid kinetic simulations of turbulence driven by the magnetorotational instability described by \cite{kunz_stone_quataert_2016} exhibit an ion distribution that is well-fit by a kappa distribution, as predicted by the GME model (Eq.~\ref{eq:fnr}), although the index appears to be harder than predicted by Eq.~\ref{eq:alphanr} for the high values of plasma beta (possibly a consequence of non-magnetic sources of free energy). Hybrid kinetic simulations of Alfv\'{e}nic turbulence may provide further tests of the model; published cases with $\delta B/B_0 \ll 1$ and moderate beta do not exihibit significant particle acceleration, consistent with Eq.~\ref{eq:alphanr} \citep[e.g.,][]{arzamasskiy_etal_2019, cerri_etal_2021}. Further benchmarking of the model in the NR regime is deferred to future work.

\section{Conclusions and discussion}

This Letter provides an analytical model for power-law nonthermal distributions that arise in collisionless plasmas due to generic energization processes. Unlike many works in the literature, this model is based on maximum entropy principles (of a generalized, non-BG form), rather than the details of the microscopic mechanisms that ultimately enable (or counteract) the acceleration. The GME distribution (Eqs.~\ref{eq:maxentropy}-\ref{eq:fnr}) provides a physically-motivated reduced model for nonthermal particle populations. Likewise, the model for the power-law index $\alpha$ of the equilibrium distribution versus plasma parameters (Eqs.~\ref{eq:alphaur}-\ref{eq:alphanr}) may be a useful prescription for systems where magnetic dissipation is the key energizer (e.g., magnetic reconnection, turbulence, and some instabilities). Further comparison to kinetic simulations will be essential for benchmarking the validity of the model and determining a more rigorous closure for $p_{c,\chi_d}/p_{c,\infty}$. Extension of the model to other processes (such as collisionless shocks) may require taking into account additional effects, such as particle escape and the self-consistent generation of magnetic fields.

There are several physical effects that may prevent the nonthermal GME state described in this Letter from being attained in some systems. First and foremost, the competition of entropy production mechanisms at multiple scales would invalidate the core assumption of the distribution being governed by $p_{c,\chi}$ at a single dominant value of the index $\chi = \chi_d$. Second, the time-dependence of physical parameters (e.g., growing plasma beta amid dissipation) may cause $p_{c,\chi_d}/p_{c,\infty}$ to vary over time, leading to hysteresis that is not accounted for in the model. These assumptions may be relaxed in future iterations of the model.

Another effect that may prevent the GME state from being attained is anisotropy of the momentum distribution (at macroscopic scales). This may occur if the energization mechanisms are strongly anisotropic with respect to the large-scale magnetic field and pitch angle scattering is inefficient. Anisotropy reduces the entropy and thus prevents complete relaxation to the (isotropic) GME state.

The GME model indicates that particle acceleration will be inefficient if the mechanisms of entropy production are localized at energy scales near the thermal energy (Landau damping being one such example). This may be the situation for simplified or dynamically constrained setups such as the collision of Alfv\'{e}n waves \citep{nattila_beloborodov_2022}, 2D NR magnetic reconnection \citep{dahlin_etal_2017, li_etal_2019}, or magnetic reconnection in a strong guide field \citep{werner_uzdensky_2017, arnold_etal_2021}.

Beyond numerical simulations, we note that {\it in situ} measurements of particle distributions in the solar wind may provide an additional test of the GME model in the NR regime. The nonthermal population of high-energy electrons (called the halo) is well fit by a kappa distribution, the parameters of which can be measured as a function of plasma conditions and distance from the Sun \citep[e.g.][]{maksimovic_etal_2005, stverak_etal_2009, abraham_etal_2022}. While the measured kappa indices at $\sim 1$ AU are reasonable in comparison to the GME model for magnetic dissipation (with $\beta \lesssim 1$ and $\delta B/B_0 \sim 1$), a careful analysis is necessary to take into account the distribution evolution from the solar corona and the possible effect of non-negligible collisions. Furthermore, the measured solar wind distribution also has a thermal component (called the core) and beamed component (called the strahl), which are not readily explained by the GME model.

Nonthermal particle acceleration is usually modeled in the language of quasilinear theory, involving concepts such as the Fokker-Planck equation (or its extensions), pitch-angle scattering, and trapping (or escape) mechanisms \citep[see, e.g.,][]{kulsrud_ferrari_1971, blandford_eichler_1987, schlickeiser_1989, chandran_2000, isliker_etal_2017, lemoine_malkov_2020, demidem_etal_2020, lemoine_2021, vega_etal_2022}. The maximum-entropy model proposed in this Letter stands in stark contrast to these conventional approaches, being only weakly dependent on the physical ingredients responsible for enabling the GME state. The two frameworks are not mutually exclusive, however, as the GME distribution may be maintained by a broad class of Fokker-Planck diffusion/advection coefficients \citep[e.g.,][]{shizgal_2018}. It is important for future work to bridge the two mathematical frameworks.

The author thanks Dmitri Uzdensky, Mitch Begelman, Greg Werner, and Yuri Levin for helpful discussions during the early stages of this project. The author is supported by a Flatiron Research Fellowship at the Flatiron Institute, Simons Foundation. Research at the Flatiron Institute is supported by the Simons Foundation.

Declaration of Interests. The author reports no conflict of interest.

\bibliographystyle{jpp}


\begin{thebibliography}{63}
\expandafter\ifx\csname natexlab\endcsname\relax\def\natexlab#1{#1}\fi
\def\au#1{#1} \def\ed#1{#1} \def\yr#1{#1}\def\at#1{#1}\def\jt#1{\textit{#1}}
  \def\bt#1{#1}\def\bvol#1{\textbf{#1}} \def\vol#1{#1} \def\pg#1{#1}
  \def\publ#1{#1}\def\arxiv#1{#1}\def\org#1{#1}\def\st#1{\textit{#1}}

\bibitem[Abraham {\em et~al.\/}(2022)Abraham, Owen, Verscharen, Bakrania,
  Stansby, Wicks, Nicolaou, Whittlesey, Agudelo~Rueda, Bercic {\em
  et~al.\/}]{abraham_etal_2022}
{\sc \au{Abraham, J.~B.}, \au{Owen, C.}, \au{Verscharen, D.}, \au{Bakrania,
  M.}, \au{Stansby, D.}, \au{Wicks, R.}, \au{Nicolaou, G.}, \au{Whittlesey,
  P.}, \au{Agudelo~Rueda, J.}, \au{Bercic, L.} \& \au{others}} \yr{2022}
  \at{Radial evolution of thermal and suprathermal electron populations in the
  slow solar wind from 0.13 to 0.5 au: Parker solar probe observations}.
  \jt{The Astrophysical Journal} .

\bibitem[Alves {\em et~al.\/}(2018)Alves, Zrake \& Fiuza]{alves_etal_2018}
{\sc \au{Alves, E.~P.}, \au{Zrake, J.} \& \au{Fiuza, F.}} \yr{2018}
  \at{Efficient nonthermal particle acceleration by the kink instability in
  relativistic jets}.  \jt{Physical review letters}  \bvol{121}~(24),
  \pg{245101}.

\bibitem[Arnold {\em et~al.\/}(2021)Arnold, Drake, Swisdak, Guo, Dahlin, Chen,
  Fleishman, Glesener, Kontar, Phan {\em et~al.\/}]{arnold_etal_2021}
{\sc \au{Arnold, H.}, \au{Drake, J.}, \au{Swisdak, M.}, \au{Guo, F.},
  \au{Dahlin, J.}, \au{Chen, B.}, \au{Fleishman, G.}, \au{Glesener, L.},
  \au{Kontar, E.}, \au{Phan, T.} \& \au{others}} \yr{2021}  \at{Electron
  acceleration during macroscale magnetic reconnection}.  \jt{Physical review
  letters}  \bvol{126}~(13),  \pg{135101}.

\bibitem[Arzamasskiy {\em et~al.\/}(2019)Arzamasskiy, Kunz, Chandran \&
  Quataert]{arzamasskiy_etal_2019}
{\sc \au{Arzamasskiy, L.}, \au{Kunz, M.~W.}, \au{Chandran, B.~D.} \&
  \au{Quataert, E.}} \yr{2019}  \at{Hybrid-kinetic simulations of ion heating
  in alfv{\'e}nic turbulence}.  \jt{The Astrophysical Journal}  \bvol{879}~(1),
   \pg{53}.

\bibitem[Aschwanden(2002)]{aschwanden_2002}
{\sc \au{Aschwanden, M.~J.}} \yr{2002}  \at{Particle acceleration and
  kinematics in solar flares--a synthesis of recent observations and
  theoretical concepts (invited review)}.  \jt{Space Science Reviews}
  \bvol{101}~(1-2),  \pg{1--227}.

\bibitem[Ball {\em et~al.\/}(2018)Ball, Sironi \& {\"O}zel]{ball_etal_2018}
{\sc \au{Ball, D.}, \au{Sironi, L.} \& \au{{\"O}zel, F.}} \yr{2018}
  \at{Electron and proton acceleration in trans-relativistic magnetic
  reconnection: dependence on plasma beta and magnetization}.  \jt{The
  Astrophysical Journal}  \bvol{862}~(1),  \pg{80}.

\bibitem[Birn {\em et~al.\/}(2012)Birn, Artemyev, Baker, Echim, Hoshino \&
  Zelenyi]{birn_etal_2012}
{\sc \au{Birn, J.}, \au{Artemyev, A.}, \au{Baker, D.}, \au{Echim, M.},
  \au{Hoshino, M.} \& \au{Zelenyi, L.}} \yr{2012}  \at{Particle acceleration in
  the magnetotail and aurora}.  \jt{Space science reviews}  \bvol{173}~(1-4),
  \pg{49--102}.

\bibitem[Blandford \& Eichler(1987)]{blandford_eichler_1987}
{\sc \au{Blandford, R.} \& \au{Eichler, D.}} \yr{1987}  \at{Particle
  acceleration at astrophysical shocks: A theory of cosmic ray origin}.
  \jt{Physics Reports}  \bvol{154}~(1),  \pg{1--75}.

\bibitem[Bulanov {\em et~al.\/}(2015)Bulanov, Esirkepov, Kando, Koga, Kondo \&
  Korn]{bulanov_etal_2015}
{\sc \au{Bulanov, S.}, \au{Esirkepov, T.~Z.}, \au{Kando, M.}, \au{Koga, J.},
  \au{Kondo, K.} \& \au{Korn, G.}} \yr{2015}  \at{On the problems of
  relativistic laboratory astrophysics and fundamental physics with super
  powerful lasers}.  \jt{Plasma Physics Reports}  \bvol{41}~(1),  \pg{1--51}.

\bibitem[Caprioli \& Spitkovsky(2014)]{caprioli_spitkovsky_2014}
{\sc \au{Caprioli, D.} \& \au{Spitkovsky, A.}} \yr{2014}  \at{Simulations of
  ion acceleration at non-relativistic shocks. i. acceleration efficiency}.
  \jt{The Astrophysical Journal}  \bvol{783}~(2),  \pg{91}.

\bibitem[Cerri {\em et~al.\/}(2021)Cerri, Arzamasskiy \& Kunz]{cerri_etal_2021}
{\sc \au{Cerri, S.~S.}, \au{Arzamasskiy, L.} \& \au{Kunz, M.~W.}} \yr{2021}
  \at{On stochastic heating and its phase-space signatures in low-beta kinetic
  turbulence}.  \jt{The Astrophysical Journal}  \bvol{916}~(2),  \pg{120}.

\bibitem[Chandran(2000)]{chandran_2000}
{\sc \au{Chandran, B.~D.}} \yr{2000}  \at{Scattering of energetic particles by
  anisotropic magnetohydrodynamic turbulence with a goldreich-sridhar power
  spectrum}.  \jt{Physical Review Letters}  \bvol{85}~(22),  \pg{4656}.

\bibitem[Comisso \& Sironi(2018)]{comisso_sironi_2018}
{\sc \au{Comisso, L.} \& \au{Sironi, L.}} \yr{2018}  \at{Particle acceleration
  in relativistic plasma turbulence}.  \jt{Physical review letters}
  \bvol{121}~(25),  \pg{255101}.

\bibitem[Comisso \& Sironi(2019)]{comisso_sironi_2019}
{\sc \au{Comisso, L.} \& \au{Sironi, L.}} \yr{2019}  \at{The interplay of
  magnetically dominated turbulence and magnetic reconnection in producing
  nonthermal particles}.  \jt{The Astrophysical Journal}  \bvol{886}~(2),
  \pg{122}.

\bibitem[Dahlin {\em et~al.\/}(2017)Dahlin, Drake \& Swisdak]{dahlin_etal_2017}
{\sc \au{Dahlin, J.}, \au{Drake, J.} \& \au{Swisdak, M.}} \yr{2017}  \at{The
  role of three-dimensional transport in driving enhanced electron acceleration
  during magnetic reconnection}.  \jt{Physics of Plasmas}  \bvol{24}~(9),
  \pg{092110}.

\bibitem[Demidem {\em et~al.\/}(2020)Demidem, Lemoine \&
  Casse]{demidem_etal_2020}
{\sc \au{Demidem, C.}, \au{Lemoine, M.} \& \au{Casse, F.}} \yr{2020}
  \at{Particle acceleration in relativistic turbulence: A theoretical
  appraisal}.  \jt{Physical Review D}  \bvol{102}~(2),  \pg{023003}.

\bibitem[Eyink(2018)]{eyink_2018}
{\sc \au{Eyink, G.~L.}} \yr{2018}  \at{Cascades and dissipative anomalies in
  nearly collisionless plasma turbulence}.  \jt{Physical Review X}
  \bvol{8}~(4),  \pg{041020}.

\bibitem[Fermi(1949)]{fermi_1949}
{\sc \au{Fermi, E.}} \yr{1949}  \at{On the origin of the cosmic radiation}.
  \jt{Physical Review}  \bvol{75}~(8),  \pg{1169}.

\bibitem[Fermi(1954)]{fermi_1954}
{\sc \au{Fermi, E.}} \yr{1954}  \at{Galactic magnetic fields and the origin of
  cosmic radiation.}  \jt{The Astrophysical Journal}  \bvol{119},  \pg{1}.

\bibitem[Fisk \& Gloeckler(2007)]{fisk_gloeckler_2007}
{\sc \au{Fisk, L.} \& \au{Gloeckler, G.}} \yr{2007}  \at{Acceleration and
  composition of solar wind suprathermal tails}.  \jt{Space science reviews}
  \bvol{130}~(1-4),  \pg{153--160}.

\bibitem[Guo {\em et~al.\/}(2014)Guo, Li, Daughton \& Liu]{guo_etal_2014}
{\sc \au{Guo, F.}, \au{Li, H.}, \au{Daughton, W.} \& \au{Liu, Y.-H.}} \yr{2014}
   \at{Formation of hard power laws in the energetic particle spectra resulting
  from relativistic magnetic reconnection}.  \jt{Physical Review Letters}
  \bvol{113},  \pg{155005}.

\bibitem[Guo {\em et~al.\/}(2021)Guo, Li, Daughton, Li, Kilian, Liu, Zhang \&
  Zhang]{guo_etal_2021}
{\sc \au{Guo, F.}, \au{Li, X.}, \au{Daughton, W.}, \au{Li, H.}, \au{Kilian,
  P.}, \au{Liu, Y.-H.}, \au{Zhang, Q.} \& \au{Zhang, H.}} \yr{2021}
  \at{Magnetic energy release, plasma dynamics, and particle acceleration in
  relativistic turbulent magnetic reconnection}.  \jt{The Astrophysical
  Journal}  \bvol{919}~(2),  \pg{111}.

\bibitem[Hankla {\em et~al.\/}(2022)Hankla, Zhdankin, Werner, Uzdensky \&
  Begelman]{hankla_etal_2022}
{\sc \au{Hankla, A.~M.}, \au{Zhdankin, V.}, \au{Werner, G.~R.}, \au{Uzdensky,
  D.~A.} \& \au{Begelman, M.~C.}} \yr{2022}  \at{Kinetic simulations of
  imbalanced turbulence in a relativistic plasma: Net flow and particle
  acceleration}.  \jt{Monthly Notices of the Royal Astronomical Society}
  \bvol{509}~(3),  \pg{3826--3841}.

\bibitem[Hoshino(2013)]{hoshino_2013}
{\sc \au{Hoshino, M.}} \yr{2013}  \at{Particle acceleration during
  magnetorotational instability in a collisionless accretion disk}.  \jt{The
  Astrophysical Journal}  \bvol{773}~(2),  \pg{118}.

\bibitem[Isliker {\em et~al.\/}(2017)Isliker, Vlahos \&
  Constantinescu]{isliker_etal_2017}
{\sc \au{Isliker, H.}, \au{Vlahos, L.} \& \au{Constantinescu, D.}} \yr{2017}
  \at{Fractional transport in strongly turbulent plasmas}.  \jt{Physical Review
  Letters}  \bvol{119}~(4),  \pg{045101}.

\bibitem[Kulsrud \& Ferrari(1971)]{kulsrud_ferrari_1971}
{\sc \au{Kulsrud, R.~M.} \& \au{Ferrari, A.}} \yr{1971}  \at{The relativistic
  quasilinear theory of particle acceleration by hydromagnetic turbulence}.
  \jt{Astrophysics and Space Science}  \bvol{12}~(2),  \pg{302--318}.

\bibitem[Kunz {\em et~al.\/}(2016)Kunz, Stone \&
  Quataert]{kunz_stone_quataert_2016}
{\sc \au{Kunz, M.~W.}, \au{Stone, J.~M.} \& \au{Quataert, E.}} \yr{2016}
  \at{Magnetorotational turbulence and dynamo in a collisionless plasma}.
  \jt{Physical Review Letters}  \bvol{117}~(23),  \pg{235101}.

\bibitem[Lemoine(2021)]{lemoine_2021}
{\sc \au{Lemoine, M.}} \yr{2021}  \at{Particle acceleration in strong mhd
  turbulence}.  \jt{Physical Review D}  \bvol{104}~(6),  \pg{063020}.

\bibitem[Lemoine \& Malkov(2020)]{lemoine_malkov_2020}
{\sc \au{Lemoine, M.} \& \au{Malkov, M.~A.}} \yr{2020}  \at{Power-law spectra
  from stochastic acceleration}.  \jt{Monthly Notices of the Royal Astronomical
  Society}  \bvol{499}~(4),  \pg{4972--4983}.

\bibitem[Leubner(2002)]{leubner_2002}
{\sc \au{Leubner, M.~P.}} \yr{2002}  \at{A nonextensive entropy approach to
  kappa-distributions}.  \jt{Astrophysics and space science}  \bvol{282}~(3),
  \pg{573--579}.

\bibitem[Ley {\em et~al.\/}(2019)Ley, Riquelme, Sironi, Verscharen \&
  Sandoval]{ley_etal_2019}
{\sc \au{Ley, F.}, \au{Riquelme, M.}, \au{Sironi, L.}, \au{Verscharen, D.} \&
  \au{Sandoval, A.}} \yr{2019}  \at{Stochastic ion acceleration by the
  ion-cyclotron instability in a growing magnetic field}.  \jt{The
  Astrophysical Journal}  \bvol{880}~(2),  \pg{100}.

\bibitem[Li {\em et~al.\/}(2019)Li, Guo, Li, Stanier \& Kilian]{li_etal_2019}
{\sc \au{Li, X.}, \au{Guo, F.}, \au{Li, H.}, \au{Stanier, A.} \& \au{Kilian,
  P.}} \yr{2019}  \at{Formation of power-law electron energy spectra in
  three-dimensional low-$\beta$ magnetic reconnection}.  \jt{The Astrophysical
  Journal}  \bvol{884}~(2),  \pg{118}.

\bibitem[Livadiotis \& McComas(2009)]{livadiotis_mccomas_2009}
{\sc \au{Livadiotis, G.} \& \au{McComas, D.}} \yr{2009}  \at{Beyond kappa
  distributions: Exploiting tsallis statistical mechanics in space plasmas}.
  \jt{Journal of Geophysical Research: Space Physics}  \bvol{114}~(A11).

\bibitem[Livadiotis \& McComas(2013)]{livadiotis_mccomas_2013}
{\sc \au{Livadiotis, G.} \& \au{McComas, D.}} \yr{2013}  \at{Understanding
  kappa distributions: A toolbox for space science and astrophysics}.
  \jt{Space Science Reviews}  \bvol{175}~(1),  \pg{183--214}.

\bibitem[Maksimovic {\em et~al.\/}(2005)Maksimovic, Zouganelis, Chaufray,
  Issautier, Scime, Littleton, Marsch, McComas, Salem, Lin {\em
  et~al.\/}]{maksimovic_etal_2005}
{\sc \au{Maksimovic, M.}, \au{Zouganelis, I.}, \au{Chaufray, J.-Y.},
  \au{Issautier, K.}, \au{Scime, E.}, \au{Littleton, J.}, \au{Marsch, E.},
  \au{McComas, D.}, \au{Salem, C.}, \au{Lin, R.} \& \au{others}} \yr{2005}
  \at{Radial evolution of the electron distribution functions in the fast solar
  wind between 0.3 and 1.5 au}.  \jt{Journal of Geophysical Research: Space
  Physics}  \bvol{110}~(A9).

\bibitem[Milovanov \& Zelenyi(2000)]{milovanov_zelenyi_2000}
{\sc \au{Milovanov, A.} \& \au{Zelenyi, L.}} \yr{2000}  \at{Functional
  background of the tsallis entropy:" coarse-grained" systems and" kappa"
  distribution functions}.  \jt{Nonlinear Processes in Geophysics}
  \bvol{7}~(3/4),  \pg{211--221}.

\bibitem[{Nalewajko} {\em et~al.\/}(2016){Nalewajko}, {Zrake}, {Yuan}, {East}
  \& {Blandford}]{nalewajko_etal_2016}
{\sc \au{{Nalewajko}, K.}, \au{{Zrake}, J.}, \au{{Yuan}, Y.}, \au{{East},
  W.~E.} \& \au{{Blandford}, R.~D.}} \yr{2016}  \at{{Kinetic Simulations of the
  Lowest-order Unstable Mode of Relativistic Magnetostatic Equilibria}}.
  \jt{The Astrophysical Journal}  \bvol{826},  \pg{115}.

\bibitem[N{\"a}ttil{\"a} \& Beloborodov(2022)]{nattila_beloborodov_2022}
{\sc \au{N{\"a}ttil{\"a}, J.} \& \au{Beloborodov, A.~M.}} \yr{2022}
  \at{Heating of magnetically dominated plasma by alfv{\'e}n-wave turbulence}.
  \jt{Physical Review Letters}  \bvol{128}~(7),  \pg{075101}.

\bibitem[Parker \& Tidman(1958)]{parker_tidman_1958}
{\sc \au{Parker, E.} \& \au{Tidman, D.}} \yr{1958}  \at{Suprathermal
  particles}.  \jt{Physical Review}  \bvol{111}~(5),  \pg{1206}.

\bibitem[Pierrard \& Lazar(2010)]{pierrard_lazar_2010}
{\sc \au{Pierrard, V.} \& \au{Lazar, M.}} \yr{2010}  \at{Kappa distributions:
  theory and applications in space plasmas}.  \jt{Solar physics}
  \bvol{267}~(1),  \pg{153--174}.

\bibitem[R{\'e}nyi(1961)]{renyi_1961}
{\sc \au{R{\'e}nyi, A.}} \yr{1961} On measures of entropy and information.
  \bt{In {\em Proceedings of the Fourth Berkeley Symposium on Mathematical
  Statistics and Probability, Volume 1: Contributions to the Theory of
  Statistics\/}},  \pg{pp. 547--561}. University of California Press.

\bibitem[Schekochihin {\em et~al.\/}(2009)Schekochihin, Cowley, Dorland,
  Hammett, Howes, Quataert \& Tatsuno]{schekochihin_etal_2009}
{\sc \au{Schekochihin, A.}, \au{Cowley, S.}, \au{Dorland, W.}, \au{Hammett,
  G.}, \au{Howes, G.}, \au{Quataert, E.} \& \au{Tatsuno, T.}} \yr{2009}
  \at{Astrophysical gyrokinetics: kinetic and fluid turbulent cascades in
  magnetized weakly collisional plasmas}.  \jt{The Astrophysical Journal
  Supplement Series}  \bvol{182}~(1),  \pg{310}.

\bibitem[Schlickeiser(1989)]{schlickeiser_1989}
{\sc \au{Schlickeiser, R.}} \yr{1989}  \at{Cosmic-ray transport and
  acceleration. i-derivation of the kinetic equation and application to cosmic
  rays in static cold media. ii-cosmic rays in moving cold media with
  application to diffusive shock wave acceleration}.  \jt{The Astrophysical
  Journal}  \bvol{336},  \pg{243--293}.

\bibitem[Schroeder {\em et~al.\/}(2021)Schroeder, Howes, Kletzing, Skiff,
  Carter, Vincena \& Dorfman]{schroeder_etal_2021}
{\sc \au{Schroeder, J.~W.}, \au{Howes, G.}, \au{Kletzing, C.}, \au{Skiff, F.},
  \au{Carter, T.}, \au{Vincena, S.} \& \au{Dorfman, S.}} \yr{2021}
  \at{Laboratory measurements of the physics of auroral electron acceleration
  by alfv{\'e}n waves}.  \jt{Nature communications}  \bvol{12}~(1),  \pg{1--9}.

\bibitem[Shizgal(2018)]{shizgal_2018}
{\sc \au{Shizgal, B.~D.}} \yr{2018}  \at{Kappa and other nonequilibrium
  distributions from the fokker-planck equation and the relationship to tsallis
  entropy}.  \jt{Physical Review E}  \bvol{97}~(5),  \pg{052144}.

\bibitem[Sironi {\em et~al.\/}(2021)Sironi, Rowan \& Narayan]{sironi_etal_2021}
{\sc \au{Sironi, L.}, \au{Rowan, M.~E.} \& \au{Narayan, R.}} \yr{2021}
  \at{Reconnection-driven particle acceleration in relativistic shear flows}.
  \jt{The Astrophysical Journal Letters}  \bvol{907}~(2),  \pg{L44}.

\bibitem[Sironi \& Spitkovsky(2010)]{sironi_spitkovsky_2010}
{\sc \au{Sironi, L.} \& \au{Spitkovsky, A.}} \yr{2010}  \at{Particle
  acceleration in relativistic magnetized collisionless electron-ion shocks}.
  \jt{The Astrophysical Journal}  \bvol{726}~(2),  \pg{75}.

\bibitem[Sironi \& Spitkovsky(2014)]{sironi_spitkovsky_2014}
{\sc \au{Sironi, L.} \& \au{Spitkovsky, A.}} \yr{2014}  \at{Relativistic
  reconnection: an efficient source of non-thermal particles}.  \jt{The
  Astrophysical Journal Letters}  \bvol{783}~(1),  \pg{L21}.

\bibitem[Spitkovsky(2008)]{spitkovsky_2008}
{\sc \au{Spitkovsky, A.}} \yr{2008}  \at{Particle acceleration in relativistic
  collisionless shocks: Fermi process at last?}  \jt{The Astrophysical Journal
  Letters}  \bvol{682}~(1),  \pg{L5}.

\bibitem[{\v{S}}tver{\'a}k {\em et~al.\/}(2009){\v{S}}tver{\'a}k, Maksimovic,
  Tr{\'a}vn{\'\i}{\v{c}}ek, Marsch, Fazakerley \& Scime]{stverak_etal_2009}
{\sc \au{{\v{S}}tver{\'a}k, {\v{S}}.}, \au{Maksimovic, M.},
  \au{Tr{\'a}vn{\'\i}{\v{c}}ek, P.~M.}, \au{Marsch, E.}, \au{Fazakerley, A.~N.}
  \& \au{Scime, E.~E.}} \yr{2009}  \at{Radial evolution of nonthermal electron
  populations in the low-latitude solar wind: Helios, cluster, and ulysses
  observations}.  \jt{Journal of Geophysical Research: Space Physics}
  \bvol{114}~(A5).

\bibitem[Tsallis(1988)]{tsallis_1988}
{\sc \au{Tsallis, C.}} \yr{1988}  \at{Possible generalization of
  boltzmann-gibbs statistics}.  \jt{Journal of statistical physics}
  \bvol{52}~(1),  \pg{479--487}.

\bibitem[Uzdensky(2022)]{uzdensky_2022}
{\sc \au{Uzdensky, D.~A.}} \yr{2022}  \at{Relativistic non-thermal particle
  acceleration in two-dimensional collisionless magnetic reconnection}.
  \jt{Journal of Plasma Physics}  \bvol{88}~(1).

\bibitem[Vega {\em et~al.\/}(2022)Vega, Boldyrev, Roytershteyn \&
  Medvedev]{vega_etal_2022}
{\sc \au{Vega, C.}, \au{Boldyrev, S.}, \au{Roytershteyn, V.} \& \au{Medvedev,
  M.}} \yr{2022}  \at{Turbulence and particle acceleration in a relativistic
  plasma}.  \jt{The Astrophysical Journal Letters}  \bvol{924}~(1),  \pg{L19}.

\bibitem[Werner \& Uzdensky(2017)]{werner_uzdensky_2017}
{\sc \au{Werner, G.~R.} \& \au{Uzdensky, D.~A.}} \yr{2017}  \at{Nonthermal
  particle acceleration in 3d relativistic magnetic reconnection in pair
  plasma}.  \jt{The Astrophysical Journal Letters}  \bvol{843}~(2),  \pg{L27}.

\bibitem[Werner \& Uzdensky(2021)]{werner_uzdensky_2021}
{\sc \au{Werner, G.~R.} \& \au{Uzdensky, D.~A.}} \yr{2021}  \at{Reconnection
  and particle acceleration in three-dimensional current sheet evolution in
  moderately magnetized astrophysical pair plasma}.  \jt{Journal of Plasma
  Physics}  \bvol{87}~(6).

\bibitem[{Werner} {\em et~al.\/}(2018){Werner}, {Uzdensky}, {Begelman},
  {Cerutti} \& {Nalewajko}]{werner_etal_2018}
{\sc \au{{Werner}, G.~R.}, \au{{Uzdensky}, D.~A.}, \au{{Begelman}, M.~C.},
  \au{{Cerutti}, B.} \& \au{{Nalewajko}, K.}} \yr{2018}  \at{{Non-thermal
  particle acceleration in collisionless relativistic electron-proton
  reconnection}}.  \jt{Monthly Notices of the Royal Astronomical Society}
  \bvol{473},  \pg{4840--4861}.

\bibitem[{Werner} {\em et~al.\/}(2016){Werner}, {Uzdensky}, {Cerutti},
  {Nalewajko} \& {Begelman}]{werner_etal_2016}
{\sc \au{{Werner}, G.~R.}, \au{{Uzdensky}, D.~A.}, \au{{Cerutti}, B.},
  \au{{Nalewajko}, K.} \& \au{{Begelman}, M.~C.}} \yr{2016}  \at{{The Extent of
  Power-law Energy Spectra in Collisionless Relativistic Magnetic Reconnection
  in Pair Plasmas}}.  \jt{The Astrophysical Journal Letters}  \bvol{816},
  \pg{L8}.

\bibitem[Wong {\em et~al.\/}(2020)Wong, Zhdankin, Uzdensky, Werner \&
  Begelman]{wong_etal_2020}
{\sc \au{Wong, K.}, \au{Zhdankin, V.}, \au{Uzdensky, D.~A.}, \au{Werner, G.~R.}
  \& \au{Begelman, M.~C.}} \yr{2020}  \at{First-principles demonstration of
  diffusive-advective particle acceleration in kinetic simulations of
  relativistic plasma turbulence}.  \jt{The Astrophysical Journal Letters}
  \bvol{893}~(1),  \pg{L7}.

\bibitem[Yoo {\em et~al.\/}(2013)Yoo, Yamada, Ji \& Myers]{yoo_etal_2013}
{\sc \au{Yoo, J.}, \au{Yamada, M.}, \au{Ji, H.} \& \au{Myers, C.~E.}} \yr{2013}
   \at{Observation of ion acceleration and heating during collisionless
  magnetic reconnection in a laboratory plasma}.  \jt{Physical review letters}
  \bvol{110}~(21),  \pg{215007}.

\bibitem[Zhdankin(2021)]{zhdankin_arxiv}
{\sc \au{Zhdankin, V.}} \yr{2021}  \at{Generalized entropy production in
  collisionless plasma flows and turbulence}.  \jt{arXiv preprint
  arXiv:2110.07025} .

\bibitem[{Zhdankin}(2021)]{zhdankin_2021}
{\sc \au{{Zhdankin}, V.}} \yr{2021}  \at{{Particle Energization in Relativistic
  Plasma Turbulence: Solenoidal versus Compressive Driving}}.  \jt{The
  Astrophysical Journal}  \bvol{922}~(2),  \pg{172}.

\bibitem[Zhdankin {\em et~al.\/}(2018)Zhdankin, Uzdensky, Werner \&
  Begelman]{zhdankin_etal_2018b}
{\sc \au{Zhdankin, V.}, \au{Uzdensky, D.~A.}, \au{Werner, G.~R.} \&
  \au{Begelman, M.~C.}} \yr{2018}  \at{System-size convergence of nonthermal
  particle acceleration in relativistic plasma turbulence}.  \jt{The
  Astrophysical Journal Letters}  \bvol{867}~(1),  \pg{L18}.

\bibitem[Zhdankin {\em et~al.\/}(2017)Zhdankin, Werner, Uzdensky \&
  Begelman]{zhdankin_etal_2017}
{\sc \au{Zhdankin, V.}, \au{Werner, G.~R.}, \au{Uzdensky, D.~A.} \&
  \au{Begelman, M.~C.}} \yr{2017}  \at{Kinetic turbulence in relativistic
  plasma: From thermal bath to nonthermal continuum}.  \jt{Phys. Rev. Lett.}
  \bvol{118},  \pg{055103}.

\end{thebibliography}

\end{document}